# *Thermal deuteron-deuteron fusion in metallic targets*

Konrad Czerski, Rakesh Dubey, Gokul Haridas Das, Sreelakshmi Thulichery, Agata Kowalska, Natalia Targosz-Ślęczka, Mathieu Valat

*Institute of Physics, University of Szczecin, 70-451 Szczecin, Poland*

konrad.czerski@usz.edu.pl

## Abstract

Deuteron-deuteron (DD) fusion reactions can be investigated at extremely low energies due to the relatively low Coulomb barrier that can be further reduced by the surrounding electrons in metallic targets. Recently, instead of an exponentially decreasing reaction yield for lowering beam energies, a yield plateau of the $^{2}H(d,p)^{3}H$ reaction occurring in a deuterated Zr target was observed at beam energies as low as 1 keV. Since protons were emitted from the near-resting center mass system, they were interpreted as products of thermal fusion occurring in ion tracks induced by the beam deuterons. Here, similar measurements have been performed on Ti and Pd targets, which have different thermal characteristics and electron screening energies. Comparison of yield plateaus observed for the studied target materials confirms the thermal spike model of DD fusion, underlying importance of the enhanced deuteron diffusion in ion tracks as well as electron screening and the threshold resonance in $^{4}He$. The experimental results also enable determination of the reaction rates and cross section at thermal energies, opening new perspectives for both astrophysical and commercial applications.

## 1. Introduction

In recent decades, deuteron-deuteron (DD) fusion reactions have been intensively studied at very low energies in various metallic environments to determine the reduction of the Coulomb barrier in a dense electron gas [1-5], which was considered as a model for the strongly coupled astrophysical plasma found in white, red and brown dwarfs or gas giants [6,7]. Moreover, in the latter case, DD fusion is expected to be an important energy source, significantly increasing the core temperature of planets such as Jupiter and Saturn [8,9]. This was supported by the finding that the exponential enhancement of the reaction cross section experimentally observed for lowering deuteron energies, described by the electron screening energy, was much stronger than predicted by theoretical models [10,11]. Only in the last years, it could be shown that this effect results due to localization of electrons by crystal lattice defects and small oxygen impurities in the irradiated metallic targets [12,13]. Investigations performed on Zr target, which strongly binds hydrogen and produces a stable $ZrD_2$ stoichiometry, delivered screening energies ranging between 100 eV for a defect free target and up to 500 eV for the Zr target implanted by small oxygen doses [13]. Whereas the lowest screening energy agrees with the theoretical value [10], the higher values arise due to crystal lattice vacancies decorated by oxygen and hydrogen, which was confirmed using electron annihilation spectroscopy [14].

Surprisingly, studies of low energy DD fusion have shown that its enhancement also results from a threshold 0+ resonance [15]. Its signature could be observed in the $^{2}H(d,p)^{3}H$ reaction based on destructive interference with well known broad resonances in $^{4}He$ [16]. In agreement with theoretical predictions, this resonance was found to decay predominantly by positron-electron pair creation, which was recently confirmed by the observation of high energy bremsstrahlung and annihilation radiation [17,18]. The direct consequence of this finding was the extension of the $^{2}H(d,p)^{3}H$ reaction measurement to the previously unattainable deuteron beam energy of 1.25 keV, using a special decelaration lens system in the “eLBRUS” accelerator at the University of Szczecin, Poland [19]. We found that the experimentally measured reaction yield no longer decreases for deuteron energies below 2.5 keV – the reaction yield remains constant down to the lowest beam energy, forming a characteristic yield plateau. The significantly higher energy of emitted protons than that expected for a direct DD fusion indicates that they are produced in the almost resting center mass system.

This effect could be explained by the emission of protons from ion tracks induced by the deuteron beam, which, due to cascade collisions with target atoms, create cylindrical areas with high temperatures, exceeding even the melting point of the target material at the beginning of the ion track evolution [19]. The averaged ion track temperature for the Zr target was estimated to be around 1000°K, which results in a strongly increased diffusion of deuterons implanted in the target and a higher number of DD collisions on meV energy scale. The absolute reaction rate measured in the energy region of the yield plateau could be then determined taking into account the increased DD cross section due to the electron screening and threshold resonance effects.

The ion track model as a collision cascade leading to a thermal spike was developed in the 1970s to explain sputtering of a target surface by low energy ions [20-22]. Later, the thermal spike model was also extended to describe interaction of swift heavy ions with matter, introducing separate temperatures for the electron gas and the lattice atoms, using electron-phonon coupling as the heat transfer mechanism [23,24]. The same model can also be applied to laser irradiation with short pulses (femtosecond/picosecond) causing crystal defects, usually supplemented with molecular dynamics studies to track the motion of atoms [25,26].

In this letter, we present new measurements of the $^{2}H(d,p)^{3}H$ reaction at very low energies using Pd and Ti targets to determine the reaction yield plateaus and compare them to that previously observed for Zr. Our main purpose is to test the ion track model of the thermal proton emission, using target materials, which are characterized by different physical parameters such as heat capacity, thermal conductivity, and electron screening energies. The experimental method presented will not only allow to determine thermal DD fusion rates, but also to find metallic materials for which its value will increase significantly.

## 2. Methods

### 2.1 Experimental setup

The experiments were performed at the eLBRUS Ultra High Vacuum Accelerator of the University of Szczecin, Poland. The facility is equipped with a differential pumping system that allows achieving a vacuum pressure of $10^{-10}$ mbar in the target chamber. A high current ion beam up to 1 mA on the target [19] is provided by an ECR ion source with a very good energy definition and stability of about 10 eV. The ion beam was magnetically analyzed using 90° double focusing magnet and focused on the target to a spot of 0.5 $cm^2$ using a system of electrostatic lenses and apertures. To increase the beam current at the lowest energies (below 4 keV per deuteron) an additional deceleration lens system in front of the target chamber was used, which enabled to operate the ion source at higher electric potential and finally reduce the ion energy on the target being at the ground potential. The ion current was monitored using two different Faraday cups and determined directly on the target. Secondary electron emission from the target increased the ion beam charge collected only by 10-15% for the entire energy range of the used deuteron beam, which was checked by positively biasing the target holder with a voltage up to 60V. A single silicon detector, with a thickness of 1000 µm and a detection area of 100 $mm^2$, situated at a backward angle of 135° was employed for measuring all charged particles emitted: protons, tritons, 3He particles and electrons/positrons, produced by the DD fusion reactions. In front of the detector, a 1 µm thick aluminum foil placed to prevent elastically scattered deuterons from entering the detector. The experimentally determined energy spectra are presented in Fig. 1.

As targets, metallic Zr, Pd and $Ti_{0.993}Pd_{0.007}$ plates of 1 mm thickness were used. They were implanted with deuterons of different beam energies to a saturation level, leading to a high stoichiometric (deuterium to metal atom) ratio of 2 for zirconium and titanium targets, which strongly bind deuterium and reduce its diffusion coefficient. In the case of palladium, the stoichiometric ratio amounted only to 0.07 due to a weak hydrogen binding, correlated with a large diffusion (see Table 1). The corresponding values could be estimated at the highest beam energies (above 15 keV) for which the contribution of the electron screening effect. could be neglected.

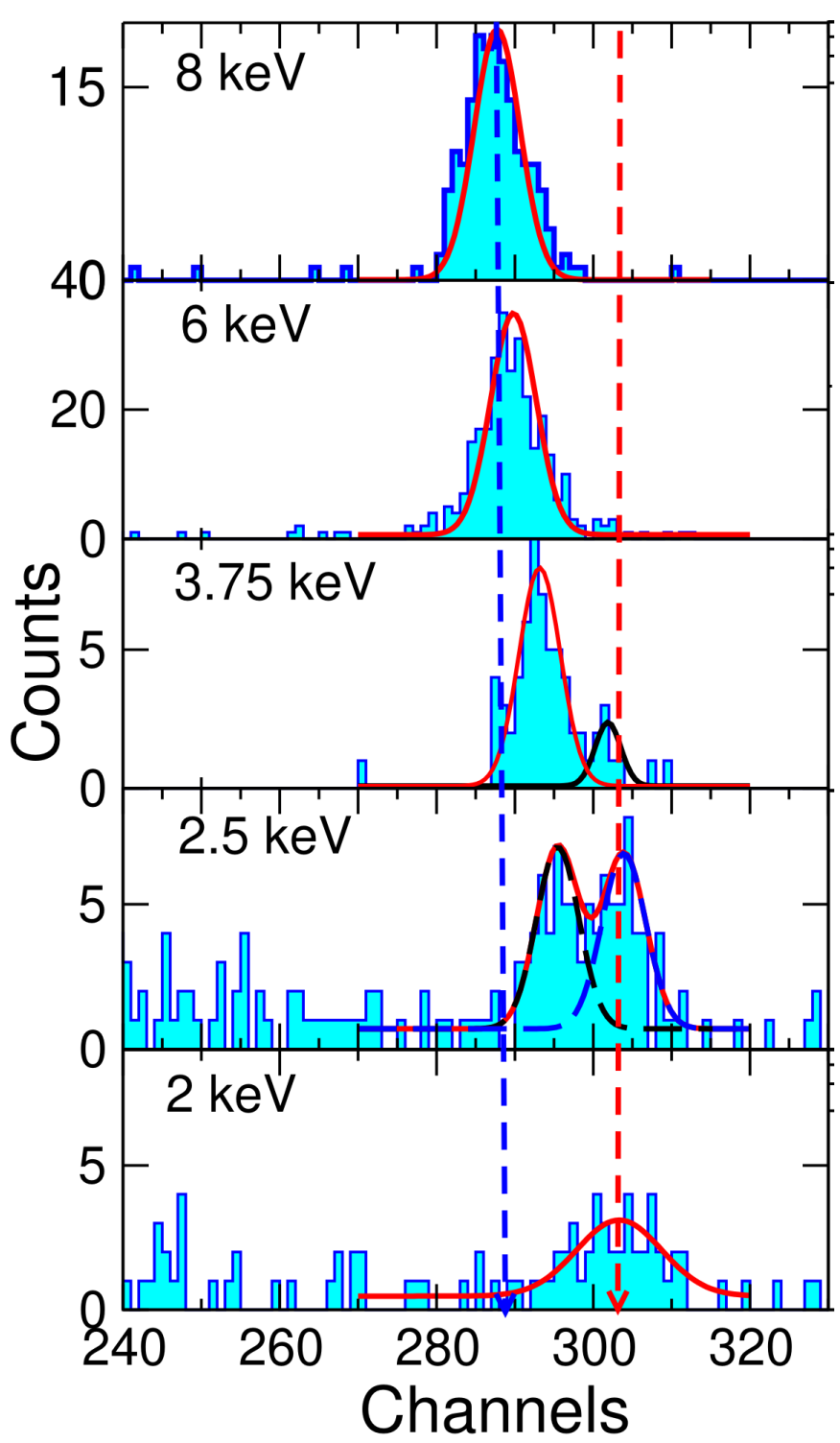

Fig. 1: Proton emission lines resulted from DD fusion in Pd, observed at different deuteron beam energies (energies are given in CMS). The channel number of the proton line related to the direct interaction with target deuterons is increasing for lowering beam energies according to the reaction kinematics – the Si-detector was placed at the backward direction of 135° with respect to the beam. The intensity of this line is strongly decreasing. The high energy line is connected to the yield plateau and to the proton emission from the resting CMS. Its position and intensity are independent of the deuteron beam energy.

### 2.2 Low energy thick target yield

Deceleration lens system has allowed to measure the reaction thick target yield measured at extremely low energies. Its value can be connected to the astrophysical S-factor using the formula:

$$Y_{scr}(E) = N_0 \int_0^R \sigma_{scr}(E)\, dx = N_0 \int_0^E \frac{\sigma_{scr}(E)}{|dE/dx|} dE \simeq \frac{2N_0\, S(E)}{C\sqrt{E_G}} \exp\left(-\sqrt{\frac{E_G}{E+U_e}}\right) \quad (1)$$

where the stopping power function in the energy region studied is governed by the electronic stopping $|dE/dx| = C\sqrt{E}$ and $N_0$ stays for the deuteron number density in the target. The expression above can be applied when the astrophysical S-factor is a weakly energy-dependent function (for details see [19]). In our case, it is a sum of two components: a constant value $S_0$ resulting from the known broad resonances in the compound nucleus $^4$He and the 0+ threshold resonance $S_R$. This resonance has a pure deuteron-deuteron structure. Therefore, the proton and neutron channels are strongly suppressed, and the electromagnetic

internal positron-electron pair creation dominates at the lowest deuteron energies. The threshold resonance S-factor can be expressed with the Bright-Wigner formula:

$$S_R(E) = \pi\sqrt{\frac{2\hbar E_G}{\mu}}\,\frac{\gamma_d^2\Gamma_f}{(E-E_R)^2+\Gamma^2(E)/4} \approx \pi\sqrt{\frac{2\hbar E_G}{\mu}}\,\frac{\gamma_d^2\Gamma_f}{E^2} \tag{2}$$

where $\Gamma(\mathrm{E})$ stays for the total widths of the threshold resonance and $\gamma_d^2 = \frac{\hbar^2}{2a\mu}$ is its single particle (deuteron) reduced widths. The total S-factor is a coherent sum of both components [15]:

$$S(E) = S_0 + S_R(E) + 2\sqrt{S_R(E)\,S_0/3}\,cos(\varphi_B^{0+}) \tag{3}$$

Finally, the enhancement factor can be then defined as the ratio between the experimentally determined reaction yield and the theoretical value determined for a fixed value of the screening energy:

$$F(E) = Y_{\mathrm{exp}}(E)/Y_{\mathrm{scr}}(E, U_e) \tag{4}$$

Most often, the enhancement factor is calculated for $U_e = 0$, but here we will use it to determine the resonance contribution to the experimental thick target yield.

**2.3 Thermal reaction rates**

At the lowest beam energies, the normalized number of detected protons does not follow “the screening curve” describing the tunneling through the screened Coulomb barrier (Eq. 1) – a constant plateau of the reaction yield is observed. As discussed in our previous paper [19], this effect can be explained with the emission of protons from the ion tracks induced by the beam in the target sample. According to the thermal spike model, in a small cylindrical region around to the primary collision cascade of impinging projectiles, the temperatures is strongly increased, even above the melting point of the studied materials. After the beam deuteron is stopped, over the next few tens of picoseconds the temperature drops and recrystallization occurs, along with the expansion of the volume of the ion track. The process can be described with the heat diffusion equation for spherical coordinates:

$$C_a\frac{\partial T(r.t)}{\partial t} = \frac{1}{r}\frac{\partial}{\partial r}[rK_a\,\partial T(r,t)/\partial r] + A_a(r,t) + gA_e(r,t) \tag{5}$$

where $C_a$ and $K_a$ are the crystal lattice heat capacity and thermal conductivity of the irradiated material, respectively. The energy absorbed within the ion track is a sum of the energy lost due to the atomic collisions and electronic excitation of the target atoms. The latter increases the temperature of the primary ion track to a fraction $g$, which arises from electron-phonon coupling for the studied material. The sum of both contributions results in the total absorbed energy in the crystal lattice $\epsilon$, which is related to the ion beam energy. The heat diffusion equation can be solved analytically for the spherical ion track and the target ambient temperature $T = 0°K$ [21]:

$$T(r,t) = \frac{\epsilon C_a^{1/2}}{(4\pi K_a t)^{3/2}} exp\left(-\frac{C_a r^2}{4K_a t}\right) \quad (6)$$

leading to a Gaussian temperature profile whose width increases with time, accompanied by a decreasing maximum value. Stopping the projectiles in the target takes about 0.1 ps, and the formation of an ion track with a well-defined temperature as a result of the collision cascade can be estimated at about 1 ps. Considering Eq.6, the temperature at the center of the ion track for the $ZrD_2$ target will reach 2980°K, and the diameter of the ion track, defined as the distance at which the temperature drops to half its central value, will be about 1 nm. For Ti and Pd targets, these values are very similar, as their heat capacities and thermal conductivities are close to those of Zr (see Table 1). Since the actual ambient temperature of ion tracks is about 300°K and the ion track expands not in the full $4\pi$ space but in half, its temperature will be considerably higher and its lifetime will be longer, which causes local melting of the target material. However, the main effect that determines the lifetime of the thermal spike, ranging between 100 and 1000 ps [25,26], is the recrystallization of the molten tracks, which generates additional heat and enhances diffusion. More accurate values require numerical modelling. Therefore, the approach presented above only illustrates which parameters of the target material determine the spatial and temporal expansion of ion tracks. Since the heat capacities and thermal conductivities and their ratio for all metallic materials studied here are very similar (see Table 1), we can conclude that the temperature profiles will be also very similar. This will also lead to similar averaged temperatures and sizes of expanding ion tracks, which are important for the enhanced diffusion of deuterium atoms, deciding about the number of deuteron-deuteron collisions and the final reaction rates.

In our previous paper [19], we proposed a simple model of ion tracks, based on the phonon density induced by projectiles in the target material due to the elastic collision cascade. It was shown that the absorbed energy density within the ion tracks remains almost constant for bombarding deuterons with energy of several keV if the electron-phonon coupling of about 10% (see Eq. 2) is assumed. This model allowed us to determine the averaged volume and temperature of the ion tracks created by the deuteron beam in a Zr target, which were about 1870nm$^3$ and 1080°K, respectively. The increased temperature leads to an enhanced deuteron diffusion and free moving deuterons in the ion tracks, which can be expressed by the Arrhenius relation:

$$N = N_0 \exp\left(-E_A/kT\right) \quad (7)$$

where $E_A$ is the activation energy and $N_0$ stays for the deuteron density in the target. Consequently, the reaction rate of the $^2$H(d,p)$^3$H reaction could be calculated assuming Maxwell-Boltzmann velocity distribution [19]:

$$\mathcal{R}(T) = N_0 N \langle \sigma\, v \rangle = N_0 N\, \frac{(8/\pi)^{1/2}}{\mu^{1/2}(kT)^{3/2}} \int_0^{\infty} \sigma(E)\, E\, exp(-E/kT)\, dE =$$
$$= N_0 N\, \frac{32\pi^{3/2}\hbar^3 E_R^{1/2}}{\mu^2\, a\, (kT)^{3/2}}\, (E_G/U_e)^{1/2}\, exp\left[-(E_G/U_e)^{1/2}\right]\, exp(-E_R/kT)\, \frac{\Gamma_p}{\Gamma} \qquad (8)$$

The formula above was derived under assumption that the single particle threshold resonance has a very small width (< 1 eV) and therefore the standard reaction rate estimation used in nuclear astrophysics for narrow resonances [27] can be applied. Accordingly, the reaction rate in the yield plateau region is determined by three most important factors: the increased deuteron diffusion determined by the density of moving deuterons $N$ , the electron screening effect given by the screening energy $U_e$ and the enhancement due to the threshold resonance described by its resonance energy and widths.

## 3. Results

### 3.1 Measurements on Ti

The experimental thick target yield determined for the Ti target using atomic and molecular deuterium beams is presented in Fig. 2. The corresponding "screening curve" was normalized to higher energy points and fitted to the data according to Eq. 1. The experimental yields measured for deuteron energies below 6 keV are significantly higher than the theoretical curve obtained with the screening energy $U_e$=25 eV (the full blue curve), which matches, however, the higher energy points very well. This yield enhancement observed for energies down to 3.5 keV could be explained with the contribution of the $0^+$ threshold resonance (the blue dotted line). We have used here the same proton resonance width of 40 meV and phase shift of 115°, which were previously obtained in our experiment on Zr [15]. Below the deuteron energy of 3.5 keV, the influence of the yield plateau is visible, value of which was estimated to be of 1.4x10$^{-4}$ (Eq. 2 and 3), being almost twice lower than for Zr [19]. Both targets were measured using the same target holder and Si detector positions so that the similar normalization factors (see Table 2) obtained in both cases with an uncertainty of about 5% indicates similar deuteron density in the target. The enhancement factor defined as a ratio between the experimental thick target yield and the theoretical value expected for the screening energy $U_e$=30 eV is additionally shown in Fig. 2. Due to the relatively small screening energy of Ti, the resonance contribution to the experimental yield can be recognized very easily.

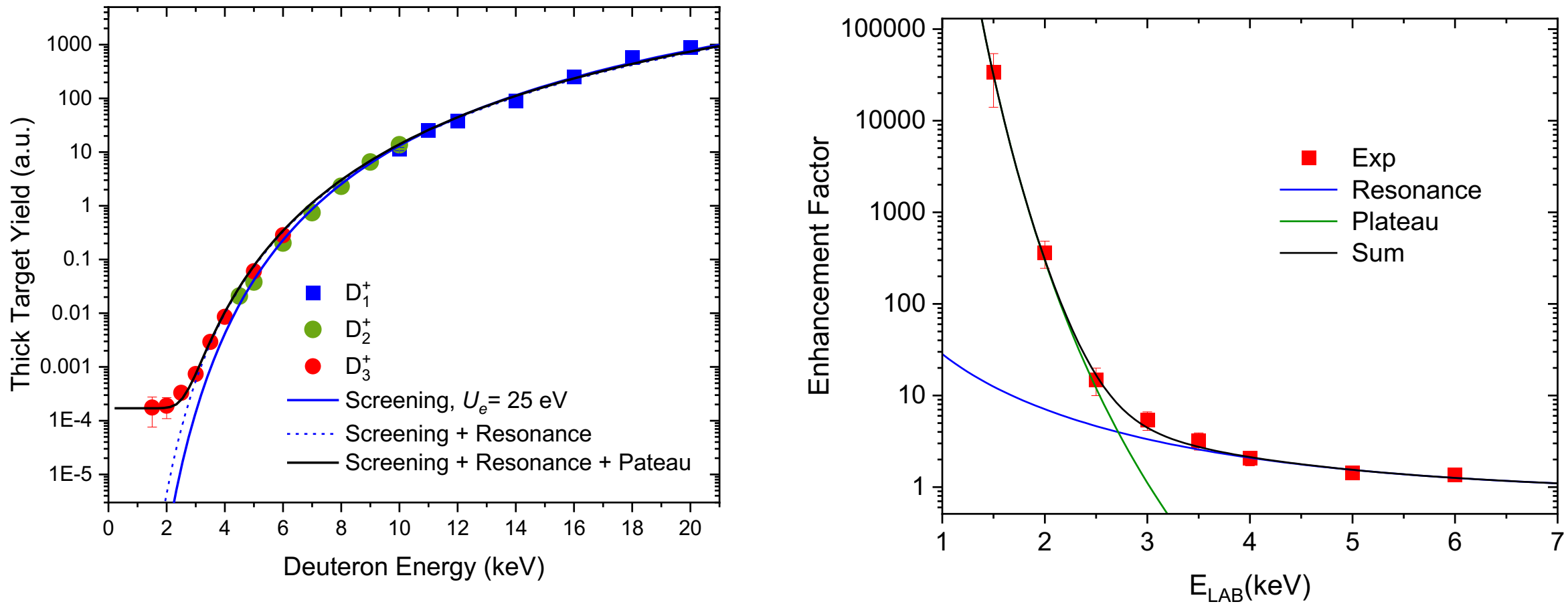

Fig. 2: Left diagram: Thick target yield of the $^2H(d,p)^3H$ reaction in Ti. The black curve represents a sum of the direct screening component and the thermal one determined by the constant yield plateau. The full blue curve represents the direct yield component calculated with $U_e$=25 eV. The dashed blue line takes into account the excitation of the threshold resonance and the black line additionally includes a constant plateau value. Right diagram: Enhancement factor determined for the $^2H(d,p)^3H$ reaction in Ti as a ratio of the measured thick target yield and the values calculated with the screening energy $U_e$=25 eV. The blue curve presents the enhancement due to threshold resonance and the green one corresponds to enhancement in the plateau energy region.

### 3.2 Measurements on Pd

The experimental data achieved for the Pd target are presented in Fig. 3. The screening energy of 400 eV was fitted without taking into account the resonance contribution, which is responsible for a much smaller yield enhancement (see the results obtained for Ti) and therefore could be neglected. The thick target yield values obtained at higher deuteron energies point to about 30 times lower deuteron density in Pd compared to that in Ti, giving the stoichiometric ratio D:M=0.05:1. At deuteron energies below 6 keV, a yield plateau is observed, the value of which is 3 orders of magnitude higher than that determined for Zr using the same normalization factor for both screening curves. For comparison, the yield plateau has been reached at deuteron energy 3 keV for Zr and 2 keV for the Ti target.

Fig. 3 also shows the position of the 3 MeV proton peak depending on the CMS energy of the deuteron beam. Since the detector is positioned at a backward angle, the energy of the emitted protons should be lower at lower projectile energies and correspond to the reaction kinematic relationship marked in black in Fig. 3. Like the results obtained before for Zr [19], we observe, however, that the protons detected in the plateau energy region are emitted from the resting frame with $E_{CM}$=0. The red curve in the right panel of Fig. 3 is calculated as a weighted average of the peak position resulting from the reaction kinematics (corresponding to the screening curve for $U_e$=400 eV in the left panel of Fig. 3) and the constant value representing the plateau yield contribution (see [19]).

At energies below 6 keV, two different spectral components of the observed proton line (see Fig. 1) can be recognized. The lower one follows the reaction dynamics, and its amplitude strongly decreases with the lowering projectile energy. The component at higher proton energy does not significantly change the count rate measured as number of counts per time and projectile current unit and can be observed at the same channel, independent of the beam energy. This component dominates the strength of the proton line in the yield plateau region and can be identify as a result of the thermal DD fusion. However, the limited resolution of the Si detector makes it very difficult to analyze the dependence of both components on the beam energy separately.

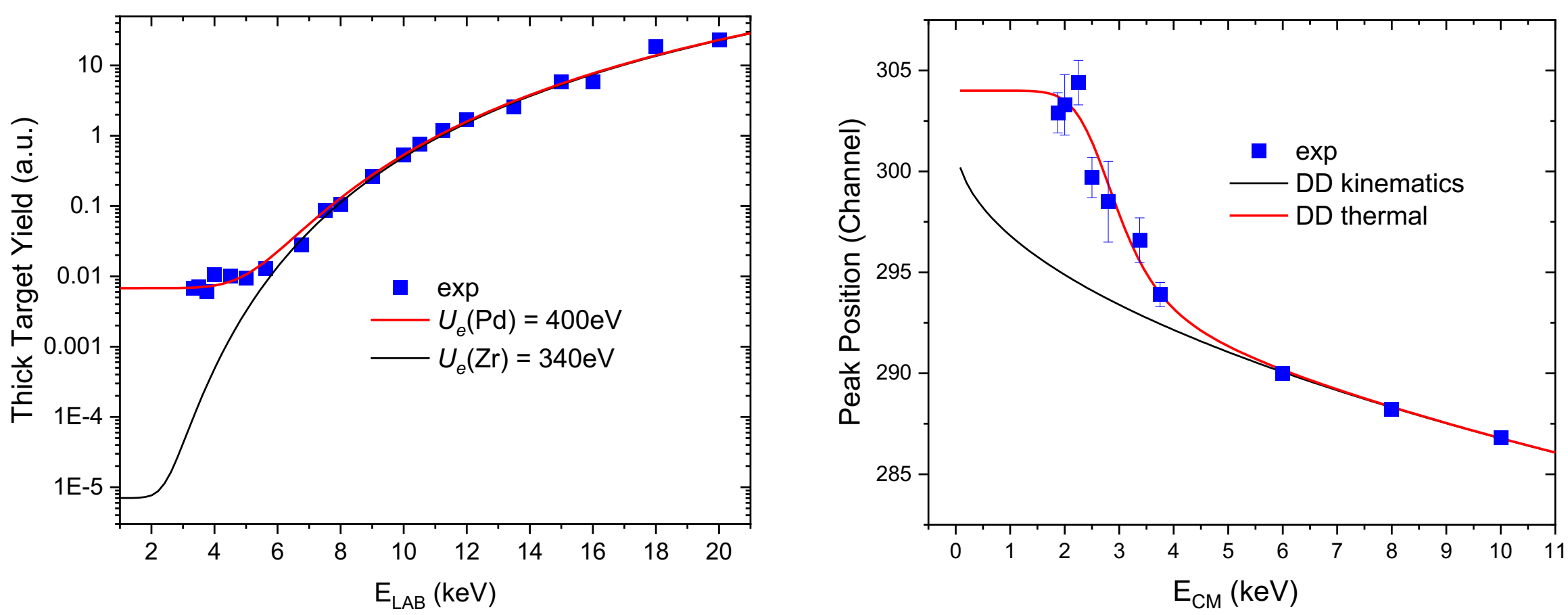

Fig. 3: Left diagram: Thick target yield of the $^{2}H(d,p)^{3}H$ reaction in Pd. The red curve represents a sum of the direct screening component and the thermal one determined by the constant yield plateau (like Fig. 2). For a comparison, the yield curve obtained for Zr, normalized to high energy points, is also presented (black curve). The yield plateau level of Pd is 3 orders of magnitude higher than that for Zr. Right diagram: Channel number of the mean proton peak position. The red curve represents a sum of both spectral contributions: direct screening component and thermal one. The proton energies measured in the yield plateau region are significantly larger than those expected due to reaction kinematics (black curve) .

### 3.3 Comparison of thermal fusion in different materials

The observation of the yield plateau and the spectral component correlated with the proton emission from the resting center mass system let us conclude that the protons are produced by DD fusion occurring at thermal energies. As already shown in our previous paper [19] this finding can be explained with ion tracks created by collision cascade induced by incident deuterons in the target. The corresponding nuclear reaction rate can be calculated using Eq. 8, which includes three different factors responsible for its strong enhancement in the ion tracks. The first factor arises from the increased lattice temperature which determines the enhanced deuterium diffusion. The second one is related to the electron screening effect and the third one to the threshold resonance. The average ion track temperature of all studied materials should be similar due to very similar ratios between heat capacity and thermal conduction coefficients. The relevant data are presented in Table 1. In our previous work [19], using a simple phonon excitation model, the average temperature of ion tracks in Zr could be estimated to be about 1080°K. Since the activation energies determining deuterium diffusion for the analyzed materials are very different, we expect that this factor can be significant for the observed yield plateau level. Especially, Pd is known for a strong hydrogen diffusion. Its activation energy of 150 meV [41] is smaller than for Zr ($E_A$=413 meV), which results in the hopping deuteron number being **16.9** times greater for Pd than for Zr. Higher screening energy leads to an additional enhancement by a factor of **61**. Both effects combined result in the 1030 times higher yield plateau, which is very close to the experimental value (see Table 2 and Fig. 3). Considering that the deuteron density was in the Pd target **37** times smaller than in Zr, we get finally the absolute plateau yield of about **27** times greater for Pd than for Zr.

Table 1. Physical parameters of studied materials.

| Target | Mass Density (g/cm$^3$) | Thermal conductivity $K_a$ (W/cm/K) | Heat Capacity $C_a$ (kJ/cm$^3$/K) | Central Temperature* (K) | Melting Point** (K) |
|---|---|---|---|---|---|
| $ZrD_{1.74}$ ($\varepsilon$) | 5.61 | 0.11[33] | 2.59[36] | 2910 | 2120 |
| $TiD_{1.93}$ ($\delta$) | 3.97 | 0.09[34] | 2.83[37] | 4110 | 1940 |
| $PdD_{0.054}$ ($\alpha$) | 12.02 | 0.10[35] | 2.93[38] | 3570 | 1830 |

* Central temperature in the ion track calculated according to Eq. 6 for t=1 ps and $\epsilon$=0.58 keV (assuming the deuteron beam energy of 2 keV [19]).
** Melting points for pure metals.

In the case of the Ti target, the measured screening energy was much lower ($U_e$=25 eV) than achieved for other metals and corresponds to values, which are expected for insulating targets. This is because of relatively thick oxygen-reach layer at the surface and embrittlement at high hydrogen density. Similar small screening energies were determined for Ti in experiments of other groups [2,3]. Additionally, due to the strong hydrogen binding in Ti, the activation energy of 680 meV [28] is even higher than for Zr, leading to 17 times weaker deuterium diffusion at 1080°K. Surprisingly, the absolute yield plateau for Ti observed experimentally is, however, only two times lower than for Zr, whereas the deuteron density is comparable (see Table 1). This finding supports the thermal spike model, which considers emission of protons from high temperature, highly excited ion track regions. As already observed in experiments using swift heavy ions [29], the irradiated insulator materials can change their structure within ion tracks to a metallic-like phase. This leads, however, to a strong increase of the screening energy, which is characteristic for metals. Consequently, the cross section for DD fusion at thermal energies rises by more than twenty orders of magnitude, allowing for observation of the yield plateau. Accordingly, the screening energy of the hot Ti target should be around 370 eV to compensate the poor hydrogen diffusion.

Table 2. Experimentally determined screening energies and yield plateaus. The activation energy from Eq. 7 describes deuterium diffusion at high temperature. The yield normalization factor corresponds is proportional to the experimental deuteron density.

| Target | D/M Ratio | Activation Energy (meV) | Yield Normalization | Screening Energy (eV) | Yield Plateau Absolute* | Yield Plateau Relative** |
|---|---|---|---|---|---|---|
| Zr | 2 | 413[39] | 14,600,000 | $340 \pm 30$ | $2.6 \times 10^{-4}$ | 1 |
| Ti | 2 | 680[40] | 12,690,000 | $25 \pm 10$ (370#) | $1.4 \times 10^{-4}$ | 0.6 |
| Pd | 0.05 | 150[41] | 390,000 | $400 \pm 40$ | $6.9 \times 10^{-3}$ | 1000 |

* Experimentally observed yield plateau.
** Yield plateau normalized on the deuteron density and the value determined for Zr.
# Screening energy for Ti, which is needed to explain the experimental yield plateau.

**4. Discussion and conclusions**

In this paper, we have presented new measurements of the $^{2}H(d,p)^{3}H$ reactions performed on deuterated Ti and Pd targets at extremely low energies. The exponential like decrease of experimental thick target yields could be described by penetration probability through the Coulomb barrier being reduced by the screening energy dependent on the target material. Like the previous experiment on Zr [19], in both cases a reaction yield plateau at the lowest deuteron energies has been observed. In particular, the results obtained for Pd are very convincing since the yield plateau could be achieved at much higher projectile energies without using a special deceleration system. Based on the proton peak energy detected by the Si-detector at backward scattering angle, it could be concluded that the DD fusion occurs in the resting center mass system, which corresponds to thermal fusion. For the first time, it has also been possible to resolve spectral contribution corresponding to the thermal and direct fusion components of the proton line, showing its different projectile energy dependence. The DD yield plateau effect could be recently confirmed also in an independent experiment [42].

All three studied metallic targets, Zr, Ti and Pd are characterized by different values of physical parameters as deuterium diffusivity, heat conduction and thermal conductivities which allows to test the ion track model of the thermal DD fusion. It assumes that the atomic collision cascade induced by projectiles locally increases crystal lattice temperature leading to enhanced deuterium diffusion and nuclear reaction rates. Formula for the nuclear reaction rates $\mathcal{R}(T)$ given by Eq. 8 makes it possible to estimate the constant plateau yield depending on the ion track temperature and its lifetime $\tau$.

$$\mathcal{N}(T) = I \cdot V \cdot \tau \cdot \frac{\Omega}{4\pi} \cdot \mathcal{R}(T) \quad (9)$$

where $V$ determines the ion track volume and $I$ stays for the deuteron beam current. The ion track parameters can be obtained from the thermal spike model (Eq. 5 and 6) using the material specific parameters: heat capacities and thermal conductivities (Table 1). The main temperature dependence results from the Arrhenius formula describing the number of colliding deuterons within ion tracks (Eq. 7). Since the uncertainties of different contributions to the reaction rate $\mathcal{R}(T)$ and material parameters are relatively large, we would like to perform here only a quantitative comparison of experiments conducted using the same measurement setup. Therefore, we assume that the average track temperature is around 1080°K for all sample materials and the ion track lifetime amounts to 100 ps. The values correspond to quantities determined for Zr in the previous work [19] utilizing a simplified thermal spike model based on phonon density induced elastic collision spike.

According to Eq. 6, the temperature at the center of the ion tracks is approximately 2900°K 1 ps after the deuteron enters the Zr target, which is also consistent with the simplified ion track model. This is the time after which the collision cascade reaches its thermodynamical balance between the electronic and lattice excitations. Since metals have high thermal conductivity coefficients, cooling of the molten ion tracks should take, in principle, a relatively short time, a few tens of ps [30]. However, the resolidification process, which adds an additional 0.1-0.2 eV per atom, extends the ion track lifetime to several hundred ps or even several fs, which was also shown in laser induced experiments [30-32]. For all studied metals, the ion track formation, due to only small differences between material parameters, should proceed very similar – in all cases the central ion track temperatures significantly exceed the melting points (see Table 1). Conservatively estimated parameter values of Eq. 9 can fully explain the observed yield plateaus in DD fusion. This can be also verified by using Eq. 8 and 9 to calculate the reaction reactivity $\langle \sigma\ v \rangle$ and cross section from the experimentally measured count rate without assuming any details about resonance and screening energy. Correspondingly, the reactivity amounts to $1.6 \times 10^{-34}$ $cm^3/s$ and the reaction cross section is about $5.3 \times 10^{-15}$ b. The latter can be estimated considering that average temperature within ion tracks is 1080°K. The result agrees very well with the previous cross section calculations based only on higher energy data [15].

One of the most important consequences of the presented experiments is also the confirmation of the existence of the threshold resonance in $^4$He. This very narrow single-particle $0^+$ resonance with the total width of about 200 meV increases the thermal reaction rate by seven orders of magnitude and was recently confirmed by observation of emission of e+e- pairs [18]. In the case of measurements on Ti, the resonance contribution could be clearly observed also at higher deuteron energies due to the very small screening energy. The enhancement factor of the $^2$H(d,p)$^3$H reaction compared to the bare nuclei cross section, neglecting the threshold resonance determined at the $E_d$=1.5 keV is larger than $10^5$.

Concluding, the experimental method employed here will likely allow to determine the reaction rates of thermal DD fusion more precisely if the material parameters will be known much more exactly and numerical calculations using for instance molecular dynamics approach will be applied. This may be particularly important for determining energy production in the interiors of brown dwarfs and giant planets. On the other hand, identifying the main factors influencing the measured reaction rates makes it easier to find special metallic materials that would produce the highest amount of energy in terrestrial laboratories.